\documentclass[%
 reprint,
groupedaddress,
showpacs,preprintnumbers,
 amsmath,amssymb,
 aps,
pra,
]{revtex4-1}

\usepackage{graphicx}
\usepackage{dcolumn}
\usepackage{bm}

\begin{document}


\title{Strong absorption of hadrons with hidden and open strangeness in nuclear matter
}

\author{
J.~Adamczewski-Musch$^{4}$, O.~Arnold$^{10,9}$, E.T.~Atomssa$^{16}$, C.~Behnke$^{8}$,
A.~Belounnas$^{16}$, A.~Belyaev$^{7}$, J.C.~Berger-Chen$^{10,9}$, J.~Biernat$^{3}$, A.~Blanco$^{1}$,
C.~~Blume$^{8}$
% ...
}

\affiliation{
\mbox{$^{1}$LIP-Laborat\'{o}rio de Instrumenta\c{c}\~{a}o e F\'{\i}sica Experimental de Part\'{\i}culas , 3004-516~Coimbra, Portugal}\\
\mbox{$^{2}$Institute of Nuclear Physics, Polish Academy of Sciences, 31342~Kraków, Poland}\\
\mbox{$^{3}$Smoluchowski Institute of Physics, Jagiellonian University of Cracow, 30-059~Krak\'{o}w, Poland}\\
\mbox{$^{4}$GSI Helmholtzzentrum f\"{u}r Schwerionenforschung GmbH, 64291~Darmstadt, Germany}\\
\mbox{$^{5}$Technische Universit\"{a}t Darmstadt, 64289~Darmstadt, Germany}\\
\mbox{$^{6}$Institut f\"{u}r Strahlenphysik, Helmholtz-Zentrum Dresden-Rossendorf, 01314~Dresden, Germany}\\
\mbox{$^{7}$Joint Institute of Nuclear Research, 141980~Dubna, Russia}\\
\mbox{$^{8}$Institut f\"{u}r Kernphysik, Goethe-Universit\"{a}t, 60438 ~Frankfurt, Germany}\\
\mbox{$^{9}$Excellence Cluster 'Origin and Structure of the Universe' , 85748~Garching, Germany}\\
\mbox{$^{10}$Physik Department E62, Technische Universit\"{a}t M\"{u}nchen, 85748~Garching, Germany}\\
\mbox{$^{11}$II.Physikalisches Institut, Justus Liebig Universit\"{a}t Giessen, 35392~Giessen, Germany}\\
\mbox{$^{12}$Institute for Nuclear Research, Russian Academy of Science, 117312~Moscow, Russia}\\
\mbox{$^{13}$Institute of Theoretical and Experimental Physics, 117218~Moscow, Russia}\\
\mbox{$^{14}$National Research Nuclear University MEPhI (Moscow Engineering Physics Institute), 115409~Moscow, Russia}\\
\mbox{$^{15}$Department of Physics, University of Cyprus, 1678~Nicosia, Cyprus}\\
\mbox{$^{16}$Institut de Physique Nucl\'{e}aire, CNRS-IN2P3, Univ. Paris-Sud, Universit\'{e} Paris-Saclay, F-91406~Orsay Cedex, France}\\
\mbox{$^{17}$Nuclear Physics Institute, The Czech Academy of Sciences, 25068~Rez, Czech Republic}\\
\mbox{$^{18}$LabCAF. F. F\'{\i}sica, Univ. de Santiago de Compostela, 15706~Santiago de Compostela, Spain}\\
\\
\mbox{$^{a}$ also at Coimbra Polytechnic - ISEC, ~Coimbra, Portugal}\\
\mbox{$^{b}$ also at ExtreMe Matter Institute EMMI, 64291~Darmstadt, Germany}\\
\mbox{$^{c}$ also at Technische Universit\"{a}t Dresden, 01062~Dresden, Germany}\\
\mbox{$^{d}$ also at Frederick University, 1036~Nicosia, Cyprus}\\
\mbox{$^{e}$ also at Dipartimento di Fisica and INFN, Universit\`{a} di Torino, 10125~Torino, Italy}\\
\mbox{$^{f}$INFN, Laboratori Nazionali di Frascati, 00044~Frascati, Italy}\\
\mbox{$^{g}$CENTRO FERMI - Museo Storico della Fisica e Centro Studi e Ricerche ``Enrico Fermi'', 00184 Rome, Italy} 
\\ $^{\dagger}$ Deceased.
}


\date{\today}

\begin{abstract}
We present the first observation of $K^-$ and $\phi$ absorption within nuclear matter by means of $\pi^-$-induced reactions on C and W targets at an incident beam momentum of $1.7~\mathrm{GeV}/c$ studied with HADES at SIS18/GSI. The double ratio $(K^{-}/K^{+})_\mathrm{W}/(K^{-}/K^{+})_\mathrm{C}$ is found to be $0.319\pm 0.009(\mathrm{stat}) ^{+ 0.014}_{-0.012}(\mathrm{syst})$ indicating a larger absorption of $K^-$ in heavier targets as compared to lighter ones. The measured $\phi/K^-$ ratios in $\pi^- + \rm C$ and $\pi^-+ \rm W$ reactions within the HADES acceptance are found to be equal to $0.55 \pm 0.03(\mathrm{stat}) ^{+ 0.06}_{-0.07}(\mathrm{syst})$ and to $0.63 \pm 0.05(\mathrm{stat}) ^{+ 0.11}_{-0.11}(\mathrm{syst})$, respectively. The similar ratios measured in the two different reactions demonstrate for the first time experimentally that the dynamics of the $\phi$ meson in nuclear medium is strongly coupled to the $K^-$ dynamics. The large difference in the $\phi$ production off C and W nuclei is discussed in terms of a strong $\phi N$ in-medium coupling.  

\end{abstract}

\maketitle

Hadron-hadron interactions are studied to understand in detail low energy QCD. The energy scale that drives these interactions is of the order of $1$ GeV and hence characterized by the non-perturbative regime, in which
effective field theories \cite{Kaiser:1995eg,Oset:1997it} can be applied. A quantitative understanding of the \mbox{(anti-)}kaon-nucleon interaction is necessary to constrain and develop these theories.

Kaon-nucleon scattering data \cite{Patrignani:2016xqp} allowed to determine the scattering parameters of the interaction that is found to be repulsive. Comparable (model-dependent) results were obtained by studying the behaviour of kaons within nuclear matter at various baryonic densities by means of (fixed target) $\gamma+A,\; p+A,\;\pi+A$ and $A+A$ reactions at kinetic energies of few GeV. Analyses of the distribution of kaon (transverse) momenta, anisotropies in the azimuthal distribution in the plane transverse to the beam direction (anisotropic flow) and cross-sections led as well to the conclusion that the real part of $K^0N$ and $K^+N$ potential is repulsive within $20-40$~MeV at saturation densities and small kaon momenta \cite{Hartnack:2011cn,Agakishiev:2014moo,Zinyuk:2014zor,Metag:2017yuh}.

The antikaon in-medium properties are more entangled \cite{Agakishiev:2009ar,Gasik:2015zwm,Adamczewski-Musch:2017rtf}. The interpretation of scattering experiments and kaonic atoms \cite{Friedman:2007zza, Bazzi:2011zj,Bazzi:2009zz,Bazzi:2010zt} is consistent with the existence of the $\Lambda(1405)$ resonance just below the $\bar{K}N$ threshold \cite{Oller:2000fj,Ikeda:2012au}. This translates on the one hand to an attractive interaction for this system and on the other hand to a large coupling of the $\bar{K}N$ channel to baryonic resonances and hence to a large imaginary part of the potential \cite{Fuchs:2005zg,Cabrera:2014lca}. To the present, absorption processes for $\bar{K}$ were studied for slow $\bar{K}$ \cite {Doce:2015ust,Piscicchia:2016dvw,Piscicchia:2018rez,Agnello:2015jfx} and hence should be extended to higher energies.

The properties of $\phi$ meson are determined by its almost pure $s\bar{s}$ content. Its narrow width ($\Gamma=4.2\; \mathrm{MeV}$) is dominated by the OZI allowed $K^+K^-$ decay \cite{PhysRevD.16.2336}.
Hence, the modification of $K^-$ spectral function are expected to have a significant impact on $\phi$ in-medium properties \cite{Cabrera:2016rnc}. The behaviour of the $\phi$ within a nuclear environment is debated since decades in a very controversial way. Indeed, in the interpretation of the ultra-relativistic heavy-ion collisions data \cite{Abelev:2014uua,Wada:2013mua} the $\phi N$ cross-sections are assumed to be
small \cite{Steinheimer:2015sha} due to OZI suppression. 
On the contrary, measurements of the modification
of the $\phi$  production rate in proton- and photon-induced
reactions off different nuclear targets do suggest a rather
sizable $\phi N$ interaction cross-section \cite{Hartmann:2012ia,Ishikawa:2004id,Muto:2005za}. These data
are, however, interpreted in a rather model-dependent
way \cite{Muhlich:2005kf,Cabrera:2017agk} implying the contribution of secondary processes, in-medium propagation, initial and final state interactions that have to be taken into account to extract final conclusions.

In this context, investigations of the $\phi$ and $K^-$ production close to production threshold in pion-induced reactions offer an ideal environment. Firstly, the production of the mesons is strongly correlated by means of two channels: (a) direct ($K^+K^-$) and (b) resonant ($\phi$) production, as known from  elementary reactions \cite{Maeda:2008prc, Balestra:2000ex,PhysRevD.16.1,Goussu:1966ps}.
Secondly, pion-induced reactions are superior to the previously studied proton- and photon-induced reactions. Due to the large $\pi N$
inelastic cross section, hadron production occurs close to the upstream surface of the nucleus, leading on average to a longer path of the produced hadrons inside the nuclear matter. Because of the much lower production cross sections of the mesons in proton-induced reactions, secondary processes are non-negligible,
causing shorter path lengths, just as in photon-induced reactions where the incident photons penetrate deeply into the nucleus.

In this work we present the first measurement of the direct coupling of the $\phi$ to antikaon properties within nuclear matter, showing that for both mesons a strong absorption is observed when comparing $\pi^-+\mathrm{W}$ to $\pi^-+\mathrm{C}$ reactions at a $\pi^-$-momentum of $1.7$ GeV/$c$.

The experiment was performed with the High Acceptance DiElectron Spectrometer (HADES) at the SIS18 at GSI Helmholtzzentrum f\"ur Schwerionenforschung in Darmstadt, Germany. HADES \cite{Agakishiev:2009am} is a charged-particle detector consisting of a six-coiled toroidal magnet surrounding the beam axis with six identical detection sections. It features polar acceptance between $18^\circ$ and $85^\circ$ with an almost full azimuthal coverage. Each sector is equipped with a Ring-Imaging Cherenkov (RICH) detector followed by Mini-Drift Chambers (MDCs), two in front of and two behind the magnetic field, as well as two detectors for time-of-flight measurements, a scintillator hodoscope (TOF) and Resistive Plate Chamber (RPC), in combination with the target-T0 detector. In the following, the TOF-RPC system is referred to as Multiplicity Electron Trigger Array (META). Hadron identification is based on time-of-flight and on specific energy-loss (d$E$/d$x$) in the MDC tracking detectors.

A secondary beam of negatively charged pions with a rate of $\approx 3\times10^5 \;\pi^-/\mathrm{s}$ and a momentum of $1.7$ GeV/$c$ was incident on carbon and tungsten targets each composed of three segments with a thickness of $3\times 7.2$ mm and $3\times 2.4$ mm, respectively. In order to measure the momentum of the secondary pion beam with a precision of $ \approx 0.3\%\,(\sigma)$, the CERBEROS tracking system \cite{Adamczewski-Musch:2017yme} was employed. The first-level-trigger (LVL1) required a signal in the target-T0 detector and a minimum charged particle multiplicity $M\geq2$ in the META system. A total of $1.3\times 10^8$ and $1.7\times 10^8$ events were collected for $\pi^-+\mathrm{C}$ and $\pi^-+\mathrm{W}$ collisions, respectively. 

\begin{figure}
\includegraphics[scale=0.45]{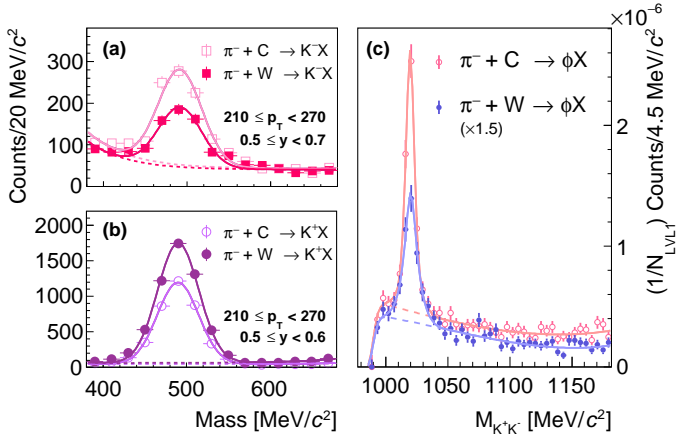}
\caption{\label{Fig:invmass_phi}(Color online) Mass distributions of $K^{-}$ (a) and $K^{+}$ (b) with the corresponding background fits (dashed lines). (c) Invariant mass distribution of $K^{+}K^{-}$ pairs normalized to the number of LVL1 events in $\pi^{-}+\mathrm{C}$ (pink points) and $\pi^{-}+\mathrm{W}$ reactions (blue points). The fit to the uncorrected experimental data consists of the sum of two Gaussians for the $\phi$ signal together with the background described by a polynominal and Gaussian function (dashed line).
}
\end{figure}

Charged kaons were pre-identified on the basis of the specific energy-loss in the MDC as a function of the momentum \citep{Wirth:2017ljg}. 
The analyses of the $\pi^- +\rm C$ and $\pi^- +\rm W$ reactions were performed in two sets of kinematic variables: $p_{T}-y$ and $p-\theta$ in the laboratory frame. The \mbox{(anti-)}kaon yield was extracted by fitting the measured mass distributions (examples in panels (a) and (b) in Fig.~\ref{Fig:invmass_phi}) obtained for the different $p_{T}-y$ ($p-\theta$) intervals \cite{Wirth:2018sqm}. The  precision of the kaon mass measurement varies between $3 - 5 \%$ and the resolution between 18.3 to 62.8 MeV/$c^2$ over the different $p_{T}-y$ bins. 
The total number of reconstructed $K^+$ and $K^-$ for the $p_{T}-y$ analysis within the HADES acceptance in $\pi^-+C$ reactions are equal to $N^{K^+}_{C} = 160,820 \pm 561$ and $N^{K^-}_{C} = 7,310 \pm 138$ and in $\pi^-+W$ reactions are equal to $N^{K^+}_{W} = 208,783 \pm 602$ and $N^{K^-}_{W} = 4,106 \pm 123$. Further, the obtained double-differential yields of $K^+$ and $K^-$ were corrected for reconstruction efficiency within the geometrical acceptance
($\approx 12 - 30 \%$ for $K^+$ and $\approx 10-26\%$ for $K^-$) and normalized to the total number of beam particles and the density of target atoms to obtain absolute cross-sections.

\begin{figure}
\includegraphics[scale=0.46]{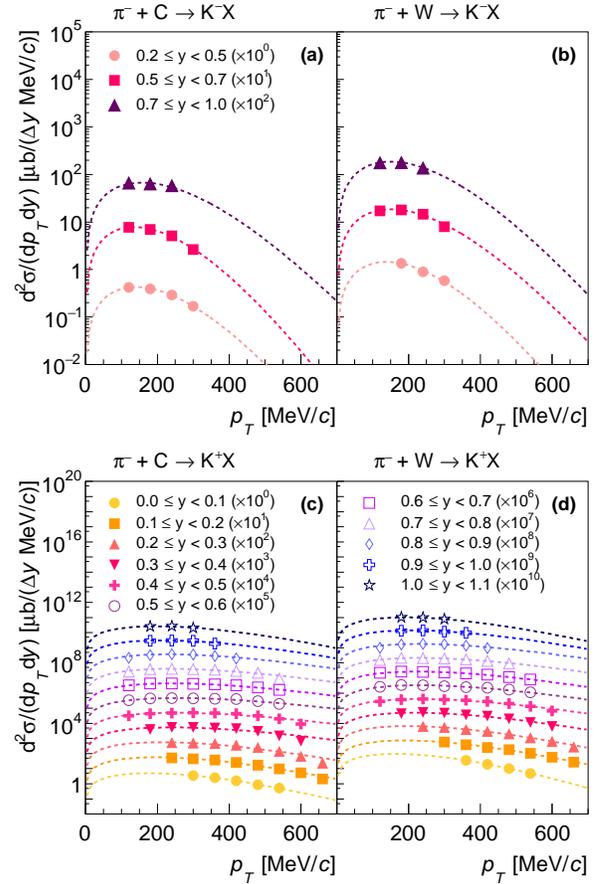}
\caption{\label{Fig:pt-y_kaon}(Color online) $K^{+}$ and $K^{-}$ differential cross-sections in different rapidity regions (see legend). The upper panel corresponds to $K^-$ and the lower panel to $K^+$. The combined, statistical and systematic, uncertainty and the normalization error are smaller than the symbol size. The dashed lines indicate Boltzmann fits (see text for details).}
\end{figure}

The resulting differential cross-sections for $K^+$ and $K^-$ produced in
$\pi^- +\rm C$ and $\pi^- +\rm W$ reactions are shown in Fig.~\ref{Fig:pt-y_kaon}. 
Panels (a) and (b) show the $p_{T}$ distributions of the $K^-$ in three different rapidity
intervals subdividing the range $0.2\leq y < 1.0$ in $\pi^-+\rm C$ and $\pi^-+\rm W$, respectively. Panel (c) depicts the $p_{T}$ distribution of the $K^+$ in 11 rapidity intervals subdividing the range $0\leq y < 1.1$. Panel (d) shows the analog for the $\pi^-+W$ system.
The errors in Fig.~\ref{Fig:pt-y_kaon} are the combined statistical and
systematic, and normalization uncertainty. The systematic uncertainty was obtained by varying the
two-dimensional \mbox{(anti-)}kaon identification cut such as to select $\approx \pm 20\%$  candidates with respect to the standard analysis. Furthermore, the width of the Gaussian function used to fit the experimental \mbox{(anti-)}kaon mass distribution is fixed to the maximum allowed limits obtained from simulations \cite{Wirth:2017ljg}. The correction uncertainty corresponds to $3\%$. The normalization error arises from the beam particle determination and is around $15\%$.
 
\begin{figure}
\includegraphics[scale=0.43]{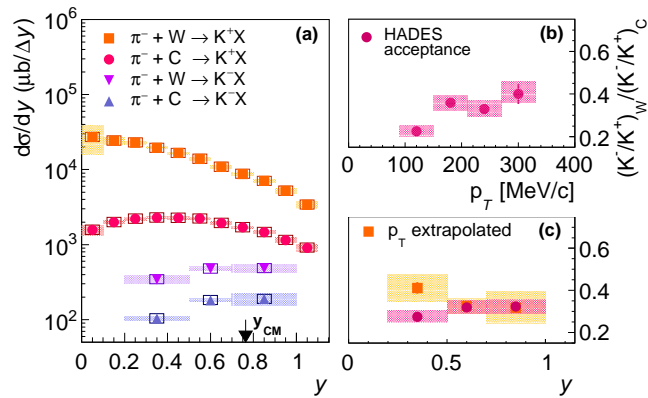}
\caption{\label{Fig:y_ratio_kaon}(Color online) (a) Cross-section of $K^+$ and  $K^-$ in $\pi^{-}+\mathrm{C}$ and $\pi^{-}+\mathrm{W}$ collisions as a function of rapidity. The shaded bands denote the systematic errors. The open boxes indicate the normalization error. The statistical uncertainties are smaller than the symbol size. The arrow ($y_{CM}$) indicates the $\pi N$ rapidity. Double ratio $(K^{-}/K^{+})_\mathrm{W}/(K^{-}/K^{+})_\mathrm{C}$ as a function of $p_{T}$ (b) and the rapidity $y$ (c) without (pink circles) and with $p_{T}$ extrapolation (orange squares) inside the HADES acceptance. The shaded areas indicate the systematic errors.}
\end{figure}

 Following the prescription already adopted in \cite{Agakishiev:2013noa,Agakishiev:2014kdy,Adamczewski-Musch:2017gjr}, a Boltzmann fit ($\mathrm{ d}^2N/(\mathrm{d}p_{T}\mathrm{d}y) = C(y)\cdot p_{T}\cdot\sqrt{p_{T}^2 + m_0^2}\cdot \exp(-\sqrt{p_{T}^2 + m_0^2}/T_{\mathrm{B}}(y))$, where $C(y)$ denotes a scaling factor, $m_0$ the nominal mass and $T_{\mathrm{B}}(y)$ the inverse-slope parameter) is applied to the measured $p_{T}$ spectra and the fit results are shown by the dashed lines in Fig.~\ref{Fig:pt-y_kaon}.
 The fits are used to extrapolate the measured distributions to uncovered transverse momentum regions. 
 The resulting experimental rapidity density distribution for $K^+$ and $K^-$ are shown in panel (a) of Fig.~\ref{Fig:y_ratio_kaon}. The extrapolation error corresponding to the uncertainty of the Boltzmann fit is the dominant contribution to the systematic uncertainty. The rapidity distribution for $K^+$ looks very different for the two colliding systems. (Elastic) scattering shifts the distribution to backward rapidity in the heavier target (W), while charge exchange is negligible \cite{Damerell:1975kw}. The shape of the $K^-$ rapidity distributions look similar for both nuclear environments, here absorption processes (e.g. $K^-N\rightarrow YN$) are dominating. 
The integrated differential production cross-section ($\Delta\sigma$) for $K^+$ ($0 \leq y < 1.1$) and $K^-$ ($0.2 \leq y < 1.0$) in $\pi^-+ \rm C$ and $\pi^-+ \rm W$ reactions inside the HADES acceptance are listed in Tab.~\ref{Tab:Xsection}.

\begin{table}
\caption{\label{Tab:Xsection}Target, particle species and cross-section inside the HADES acceptance. Error values shown are statistic (first), systematic (second) and normalization (third).}

\begin{ruledtabular}
\begin{tabular}{c c c}

Target & Particle & $\Delta\sigma$ [$\mu\mathrm{b}$] \\ \hline

C & $K^{+}$ & $1974 \mbox{$\;\pm\;$} 7 ^{+67}_{-69} \pm 310 \; \; \;$\\
C & $K^{-}$ & $124 \mbox{$\;\pm\;$} 2 \pm 11 \pm 20$ \\
C & $\phi$ & $41 \mbox{$\;\pm\;$} 2 \pm 2 ^{+6}_{-5}\;\;\;\;$ \\
W & $K^{+}$ & $15965 \mbox{$\;\pm\;$}46  ^{+1236}_{-1247} \pm 2509$ \\
W & $K^{-}$ & $345 \mbox{$\;\pm\;$} 10 \pm 26 \pm 54$ \\
W & $\phi$ & $112 \mbox{$\;\pm\;$} 8 \pm 5^{+18}_{-14}\;\;\;\;\;\,\,$ \\

\end{tabular}
\end{ruledtabular}
\end{table}

In order to study the $K^-$ absorption inside the nuclear medium, a comparison of the $K^-/K^+$ ratio measured in collisions with heavy target (W) and lighter one (C) is done. In the double ratio $(K^{-}/K^{+})_\mathrm{W}/(K^{-}/K^{+})_\mathrm{C}$ the $K^+$ acts as a reference particle for the strange hadron production, because of its very low absorption cross-section. Therefore, the measured $K^-$ distribution is normalized to the integrated $K^+$ yield ($0 \leq y < 1.1$).

The double ratio  $(K^{-}/K^{+})_\mathrm{W}/(K^{-}/K^{+})_\mathrm{C}$ inside the HADES acceptance is moderately increasing with increasing $p_{T}$ as shown in panel (b) of Fig.~\ref{Fig:y_ratio_kaon}.
Within errors no significant dependence on the rapidity $y$ (panel (c) of  Fig.~\ref{Fig:y_ratio_kaon}), with and without $p_{T}$ extrapolation, is observed with an average value of $(K^{-}/K^{+})_\mathrm{W}/(K^{-}/K^{+})_\mathrm{C} = 0.319\pm 0.009(\mathrm{stat}) ^{+ 0.014}_{-0.012}(\mathrm{syst})$ in the latter case.
In order to define a reference value in which no nuclear absorption is present, a double ratio $(K^{-}/K^{+})_\mathrm{W}/(K^{-}/K^{+})_\mathrm{C}$ was constructed from elementary $\pi^- N$ reactions by taking into account the proper number of neutrons and protons in tungsten and carbon. This procedure results in a value of $0.93\pm0.09$ \cite{Wirth:2018sqm}.
The fact that the measured double ratios are well below this reference demonstrates the $K^-$ absorption.

\begin{table}
\caption{\label{Tab:Ratios} Particle ratios inside the HADES acceptance. Error values shown are statistic (first), systematic (second) and normalization (third).}

\begin{ruledtabular}
\begin{tabular}{c c c}

Ratios &  \\ \hline

$(K^{-}/K^{+})_\mathrm{W}/(K^{-}/K^{+})_\mathrm{C}$ &  $0.319\pm 0.009 ^{+ 0.014}_{-0.012}$ \;\; \;\;\;\;\, \\
$(\phi/K^-)_{\mathrm{C}}$ &  $0.55 \pm 0.03 ^{+ 0.06}_{-0.07}$ \;\; \;\;\; \;\; \\
$(\phi/K^-)_{\mathrm{W}}$ &  $0.63 \pm 0.05 \pm 0.11$ \;\; \;\;\;\\
$T=\frac{12}{184}\frac{\Delta\sigma_{\rm W}^{ \phi}}{\Delta\sigma_{\rm C}^{\phi}}$ &  $0.18 \pm 0.02 \pm 0.01 ^{+ 0.04}_{-0.03}$ \\

\end{tabular}
\end{ruledtabular}
\end{table}

The $\phi$ mesons were identified via their dominant decay channel into $K^+K^-$ pairs 
($BR=48.9\pm0.5\%$ \cite{Patrignani:2016xqp}). Both charged kaons were selected by a $\beta$ 
vs. momentum cut ($p/\sqrt{p^2 + m_{0}^2} \pm 0.5 \gtrless \beta$, $m_{0}$ = 493.677 
MeV/$c^{2}$ \cite{Patrignani:2016xqp}). The contamination from other particle species was 
further reduced by selecting a reconstructed \mbox{(anti-)}kaon mass interval of $400 <m <600 
\;\mathrm{MeV/}c^2$. Further, the nominal mass $m_{0}$ was attributed to the identified 
\mbox{(anti-)}kaon candidates. The resulting $K^+K^-$ invariant mass distribution  for $\pi^- + \rm 
C$ and $\pi^- + \rm W$ collisions normalized to the number of LVL1 events is shown in 
panel (c) of Fig.~\ref{Fig:invmass_phi}. A clear $\phi$ peak is visible and the signal can be described by 
the sum of two Gaussian distributions to account for finite resolution effects as well as for the re-scattering of the $K^+$ and $K^-$ inside the targets. The 
background is modeled by a third-order polynomial together with a Gaussian to account for the mass threshold ($2\times m_{0}$).

The precision of the $\phi$ mass measurement is better than 1 MeV and the $\phi$ mass resolutions are $4.0\pm0.8\;\mathrm{MeV}/c^2$ and $5.5\pm2.4\;\mathrm{MeV}/c^2$ for C and W targets, respectively.
In total $N^{\phi}_{\mathrm{C}}=578\pm35$ and $N^{\phi}_{\mathrm{W}}=341\pm32$ $\phi$ mesons are reconstructed.
The acceptance and efficiency was determined by 
simulating the reaction $\pi^- + p(^{12}\mathrm{C})\rightarrow 
\phi[\rightarrow K^+K^-]+ n(^{11}\mathrm{B})$ and values of $\approx (2-15)\%$ 
were obtained. As these 
corrections strongly depend on the $\phi$ kinematics, a differential 
correction as a function of  $p_{T}-y$ ($p-\theta$) was 
evaluated \cite{Wirth:2018bqp}. Since the limited statistics does not allow for a double 
differential analysis, each $\phi$ candidate is weighted 
with its corresponding correction factor and an integrated  $K^+K^-$ 
invariant mass spectrum is built. 
The same fitting procedure employed for the uncorrected $K^+K^-$ invariant 
mass distribution to extract the raw $\phi$ yields was applied also to 
the corrected spectra. 
After the correction for the branching ratio the integrated differential $\phi$ production cross-sections ($\Delta\sigma$) within the HADES acceptance ($0.4 \leq y < 1.0$ and $150 \leq p_{T} < 650\;\mathrm{MeV}/c$) for $\pi^- + \rm C$ and $\pi^- +\rm W$ are listed in Tab.~\ref{Tab:Xsection}. The systematic uncertainty was evaluated by varying the invariant mass binning with respect to the standard analysis as well as by changing the order of the polynominal used to fit the background from third to a second order. The correction uncertainty corresponds to $3\%$ for each kaon.

The $\phi$ absorption can be quantified in terms of the transparency ratio defined as $T=\frac{12}{184}\frac{\Delta\sigma_{\rm W}^{ \phi}}{\Delta\sigma_{\rm C}^{\phi}}$. This value is equal to $0.18 \pm 0.02(\mathrm{stat}) \pm 0.01(\mathrm{syst})^{+ 0.04}_{-0.03}(\mathrm{norm})$ and hence smaller than the results obtained in $p+A$ by ANKE \cite{Polyanskiy:2010tj} and $\gamma+A$ by CLAS \cite{Wood:2010ei} for slightly bigger nuclei ($T_{ANKE}=0.29\pm0.01(\mathrm{stat})\pm0.02(\mathrm{syst})$, $T_{CLAS}=0.46\pm0.12(\mathrm{stat})\pm0.13(\mathrm{syst}$). The lower transparency ratio found in pion-induced reactions maybe attributed
to the already mentioned large $\pi N$ reaction cross-section and the negligible contribution of secondary reactions that leads to $\phi$ production near the upstream surface, allowing the hadron to travel a longer path within the nucleus than in proton- and photon-induced reactions.

The $\phi/K^-$ ratio was evaluated for both collision systems inside the HADES acceptance. For this investigation it has to be considered
that the $K^-$ and $\phi$ phase space coverage of HADES is different ($0.4 \leq y_{\phi} <~1$ and $0.2 \leq y_{K^-} < 1$, $150 \leq p_{T,\phi} <650\; \mathrm{MeV/}c$ and $90 \leq p_{T,K^-} <330 \;\mathrm{MeV/}c$). As shown in panel (a) of Fig.~\ref{Fig:y_ratio_kaon} the shape of the $K^-$ rapidity distributions do not depend on the target.
To demonstrate that also the $\phi$ distributions share the same feature, GiBUU simulations were carried out for the $\pi^-+\rm C( \rm W)$ systems and no bias due to the HADES geometrical acceptance was found.
 
The measured $\phi/K^-$ ratio within the HADES acceptance is $0.55 \pm 0.03(\mathrm{stat}) ^{+ 0.06}_{-0.07}(\mathrm{syst})$ for $\pi^- + \rm C$ and $0.63 \pm 0.05(\mathrm{stat}) ^{+ 0.11}_{-0.11}(\mathrm{syst})$ for $\pi^-+ \rm W$ collisions.
The main systematic error arises from the difference in the $p_{T}-y$ and $p-\theta$ analyses.
Within errors the two $\phi /K^- $ ratios are in agreement.
Since the double ratio $(K^{-}/K^{+})_\mathrm{W}/(K^{-}/K^{+})_\mathrm{C}$ (panel (b) in Fig.~\ref{Fig:y_ratio_kaon}) clearly indicates a larger $K^-$ absorption in the W target and since the $\phi /K^- $ ratios are the same for both targets, also a stronger $\phi$ absorption in W is observed. This demonstrates that both resonant and non-resonant channels are affected in the medium in the same way.

In summary, the inclusive production cross-sections of charged kaons and $\phi$ mesons in $\pi^- +A$ collisions within the HADES acceptance were measured. The rapidity density distributions for $K^+$ and $K^-$ produced on heavy (W) and light (C) nuclei were compared. Strong scattering effects are observed shifting the maximum of the $K^+$ distribution to backward rapidity in the heavier target, while the shape of the $K^-$ distribution is comparable in both targets. The measured double ratio $(K^{-}/K^{+})_\mathrm{W}/(K^{-}/K^{+})_\mathrm{C}$ given in Tab.~\ref{Tab:Ratios} is well below the expected \mbox{(anti-)}kaon production reference based on elementary $\pi N$ reactions, directly indicating sizable $K^-$ absorption in heavy nuclei (W) with respect to light ones (C). The $\phi/K^-$ ratios for $\pi^- + \rm C$ and $\pi^-+\rm W$ reactions within the HADES acceptance listed in in Tab.~\ref{Tab:Ratios} are consistent within the uncertainties pointing to a non-negligible $\phi$ absorption. This finding is in line with the extracted $\phi$ transparency ratio of $\approx 18\%$, which is lower than observed by ANKE \cite{Polyanskiy:2010tj} and CLAS \cite{Wood:2010ei} measurements. This first measurement of kaons and $\phi$
in the same reactions provides experimental evidence of the strong coupling
between the $\phi$ and $K^-$ dynamics within nuclear matter.

The HADES Collaboration gratefully acknowledges the support given by the following institutions and agencies:
SIP JUC Cracow, Cracow (Poland), 2017/26/M/ST2/00600; TU Darmstadt, Darmstadt (Germany), ExtreMe Matter Institute EMMI at GSI Darmstadt; Goethe-University, Frankfurt (Germany), ExtreMe Matter Institute EMMI at GSI Darmstadt; TU M\"unchen, Garching (Germany), MLL M\"unchen, DFG EClust 153, GSI TMLRG1316F, BmBF 05P15WOFCA, SFB 1258, DFG FAB898/2-2; NRNU MEPhI Moscow, Moscow (Russia), in framework of Russian Academic Excellence Project 02.a03.21.0005, Ministry of Science and Education of the Russian Federation 3.3380.2017/4.6; JLU Giessen, Giessen (Germany), BMBF:05P12RGGHM; IPN Orsay, Orsay Cedex (France), CNRS/IN2P3; NPI CAS, Rez, Rez (Czech Republic), MSMT LM2015049, OP VVV CZ.02.1.01/0.0/0.0/16 013/0001677, LTT17003.

\bibliography{apssamp}

\end{document}